\documentclass{aastex61}
\usepackage[utf8]{inputenc}
\usepackage{amsmath}
\usepackage{longtable}
\usepackage{graphicx}

\shortauthors{Oelkers \& Stassun}

\begin{document}

\title{Precision Light Curves from \textit{TESS} Full-Frame Images: A Difference Imaging Approach}

\correspondingauthor{Ryan J. Oelkers}
\email{ryan.j.oelkers@vanderbilt.edu}

\author{Ryan J. Oelkers}
\affil{Vanderbilt University, Department of Physics and Astronomy, Nashville, TN 37235}
\affil{Vanderbilt Data Science Fellow}
\affil{Vanderbilt Initiative in Data-Intensive Astrophysics (VIDA)}

\author{Keivan G. Stassun}
\affiliation{Vanderbilt University, Department of Physics and Astronomy, Nashville, TN 37235}
\affil{Vanderbilt Initiative in Data-Intensive Astrophysics (VIDA)}
\affiliation{Fisk University, Department of Physics, Nashville, TN 37208}

\begin{abstract}
The Transiting Exoplanet Survey Satellite (\textit{TESS}) will observe $\sim$150~million stars brighter than $T_{\rm mag} \approx 16$, with photometric precision from 60~ppm to 3~percent, enabling an array of exoplanet and stellar astrophysics investigations.While light curves will be provided for $\sim$400,000 targets observed at 2-min cadence, observations of most stars will only be provided as full-frame images (FFIs) at 30~min cadence. The \textit{TESS} image scale of $\sim21$''/pix is highly susceptible to crowding, blending, and source confusion, and the highly spatially variable point spread function (PSF) will challenge traditional techniques, such as aperture and Gaussian-kernel PSF photometry. We use official ``End-to-End~6" \textit{TESS} simulated FFIs to demonstrate a difference image analysis pipeline, using a $\delta$-function kernel,that achieves the mission specification noise floor of 60~ppm~hr$^{-1/2}$. We show that the pipeline performance does not depend on position across the field, and only $\sim$2\% of stars appear to exhibit residual systematics at the level of $\sim$5~ppt. We also demonstrate recoverability of planet transits, eclipsing binaries, and other variables. We provide the pipeline as an open-source tool at \url{https://github.com/ryanoelkers/DIA} in both IDL and PYTHON. We intend to extract light curves for all point sources in the \textit{TESS} FFIs as soon as they become publicly available, and will provide the light curves through the \textit{Filtergraph} data visualization service. An example data portal based on the simulated FFIs is available for inspection at \url{https://filtergraph.com/tess_ffi}. 
\end{abstract}

\section{Introduction\label{sec:intro}}

The dramatic increase in efforts to search for transiting exoplanets at the turn of the millennium motivated the use of small-aperture, wide-field surveys to simultaneously monitor as many stars as possible \citep[e.g.,][]{Pollacco2006,Pepper:2003,Bakos2002}. Coupled with technological and computing advances, which provided pathways to rapid reduction of large data sets, the resulting massive influx of time-series photometry has also helped to guide astronomy into an era of ``big-data". The next wave of large astronomical surveys is expected to further drive the astronomy community's need for powerful data reduction tools capable of providing large-scale data products with high precision and on a rapid timescale.

Once such upcoming survey is the Transiting Exoplanet Survey Satellite \citep[\textit{TESS};][]{Ricker:2014}, which will be conducting a nearly all-sky photometric survey for two years, with a core mission goal to discover small transiting exoplanets orbiting nearby bright stars. One major consequence of the \textit{TESS} survey will be the enormous amount of time-series data collected at high cadence and for a very large number of stars. While only the $\approx$400,000 highest priority transiting-planet host stars will be observed with 2-min cadence, nearly every star outside $\pm$6~deg of the ecliptic plane will be observed with at least 30-min cadence---a total of $\sim423,000,000$ stars in the TESS Input Catalog \citep[TIC;][]{Stassun:2017}---whose data will be released not as extracted light curves but in the form of full-frame images \citep[hereafter, FFIs;][]{Sullivan:2015}. 

\textit{TESS} is equipped with four 10-cm telescopes, each containing four charged couple devices (CCDs), with a total field of view of 24~deg$^2$ on the celestial sphere during a single pointing. Each telescope is aligned to cover different areas of a longitudinal strip (called a sector), thus the 4 CCDs together in a given camera observe a total coverage of $96^{\circ}\times24^{\circ}$. The satellite will observe each ecliptic hemisphere with 13 sectors over the course of a year, each sector observed continuously for 27~d. Additionally, the increasing areal overlap of sectors at higher ecliptic latitudes will combine in a way to create continuous viewing zones around the ecliptic poles, designed to: (1) match the James Webb Space Telescope continuous viewing zone, to facilitate potential follow up; and (2) avoid the ecliptic plane to minimize contamination from solar system objects. Of particular consequence for handling the FFIs, the telescope is capable of observing such large areas of the sky because of the optical design, which maps a large 21~arcsec$^2$ area onto a single \textit{TESS} pixel \citep{Ricker:2014}. 

The extraction of light curves from wide-field images with such large pixel scales is highly non-trivial. The effects of crowding and blending hinder methods such as aperture photometry, because contamination from nearby stars cannot be easily removed from even a small photometric aperture \citep{Wang2011}. Similarly, the point spread functions (PSFs) of such large fields of view are notoriously highly spatially variable, which makes photometric extraction based on PSF-fitting extremely laborious, as individual PSFs need to be constructed for different parts of the image---particularly near the edges of the frame. Wide-field images are also typically under-sampled, which has the tendency of causing the measured flux of a given star to be correlated with its position on the frame. This can introduce systematics in the light curves which are difficult to interpret and correct, and can even mimic some astrophysical signals of interest, leading to false positives \citep{AlardLupton,Alard2000,Miller2008}.  

Light curve extraction via difference imaging analysis (hereafter, DIA) is one approach used by some wide-field surveys because it mitigates many of the effects of crowded fields \citep{AlardLupton,Alard2000,Miller2008,Oelkers2015}. DIA involves the subtraction of two frames: one frame, typically the one with the higher signal-to-noise ratio (SNR), is blurred to appropriately match the seeing conditions of the other frame; then the two are subtracted to reveal any pixels whose flux has changed between the two frames. Stars with no inherent astrophysical variation in their flux will subtract to their Poisson noise level, while stars with true variations will leave a significant residual on the frame involving multiple correlated pixels within the PSF. DIA is useful in crowded environments because the majority of the stars on a given frame should subtract cleanly, and then simple aperture or PSF photometry can extract the residuals from the differenced frame without contamination from neighboring stars. 

Such an approach has already proven successful with ground-based surveys such as the Kilodegree Extremely Little Telescope (KELT), whose imaging field of view and pixel scale are very similar to that of \textit{TESS} \citep{Pepper2007,Pepper:2012}. Indeed, KELT light curves extracted via a customized DIA-based pipeline \citep{Siverd:2012} typically achieve a precision of $\lesssim$10~mmag \citep[e.g.,][]{Oelkers:2018}. 

However, most DIA pipelines, including that used by KELT, utilize a Gaussian basis to solve for the kernel that is used to blur the reference frame before subtraction. These kernels are advantageous because they require a solution for only a small number of basis vectors, and in general, the PSFs of many astronomical images are approximately Gaussian in shape. However, when the PSF of the image deviates significantly from Gaussian, as in very wide-field images like the \textit{TESS} FFIs, these methods tend to produce low-quality subtractions and begin to suffer from the similar problems of standard PSF photometry when the model PSF is poorly matched to the true PSF \citep[see, e.g.,][]{Miller2008,Oelkers2015}. 

This Letter describes an alternative DIA approach, which uses a Dirac $\delta$-function kernel to solve for non-Gaussian, arbitrarily-shaped PSFs. It is a heavily modified version of the pipeline from \citet{Oelkers2015}, optimized for \textit{TESS} FFIs. We apply the adapted pipeline to official NASA ``End-to-End~6" \citep[Tenebaum et al. 2018, in preparation, hereafter, ETE-6]{Jenkins:2004,Bryson:2010} \textit{TESS} simulated FFIs to analyze its performance relative to nominal \textit{TESS} specifications. 

The remainder of this Letter is organized as follows: \S~\ref{sec:data} describes the simulated data; \S~\ref{sec:pipeline} details our data reduction pipeline; \S~\ref{sec:testing} presents the performance of the pipeline in terms of photometric precision relative to the expected \textit{TESS} noise model; and \S~\ref{sec:summary} summarizes our results, including intended improvements to the pipeline prior to the first official \textit{TESS} data release near the end of 2018.

\section{Data: Simulated \textit{TESS} Full Frame Images \label{sec:data}}

We tested our pipeline using the ETE-6 FFIs provided by NASA \citep[Tenebaum et al. 2018, in preparation]{Jenkins:2004,Bryson:2010}. These images were built using the expected photometric precision of the satellite, PSF shapes from lab testing (with typical full-width at half-maximum of 1.88 pixels), and were designed to have a cadence similar to that expected for the mission. The location and intensity of stars on each frame were modeled after real stars found in the \textit{TESS} Input Catalog \citep[TIC]{Stassun:2017}, and several hundred of the stars were injected with signals mimicking real astrophysical variations observed by the \textit{Kepler} satellite \citep{Borucki2010}. Spacecraft jitter was also introduced into the data to simulate systematics not yet included in the official noise model (see \citealt{Sullivan:2015}, updated by Pepper et al.~(2018, in preparation), but which are expected to occur during the mission.

NASA released the ETE-6 FFIs on the Mikulski Archive for Space Telescopes (MAST) on 2018 February 20 in both uncalibrated and calibrated form. Currently, the FFIs are available for all 16 CCDs and simulate a single observing sector for 27~d. To simplify the testing process, we selected the calibrated FFIs from a single CCD camera (in this case, CCD~\#2 on Camera~2, centered on $\alpha\sim16.1$, $\delta\sim+28$). 

The reduction of all 1,348 images completed in $\sim$4.5~d and 106,399 light curves were extracted with \textit{TESS} magnitudes $7<T<16$. The tested pipeline can later be applied to all 16 CCDs in parallel, thus we expect to be able to extract light curves from the eventual real FFI stars of a given sector within $\sim$1~week of their public release.

\section{Methods: A Difference Imaging Pipeline for Wide-Field Images \label{sec:pipeline}}

\begin{figure}[!ht]
    \centering
    \includegraphics[width=\textwidth,height=\textheight,keepaspectratio]{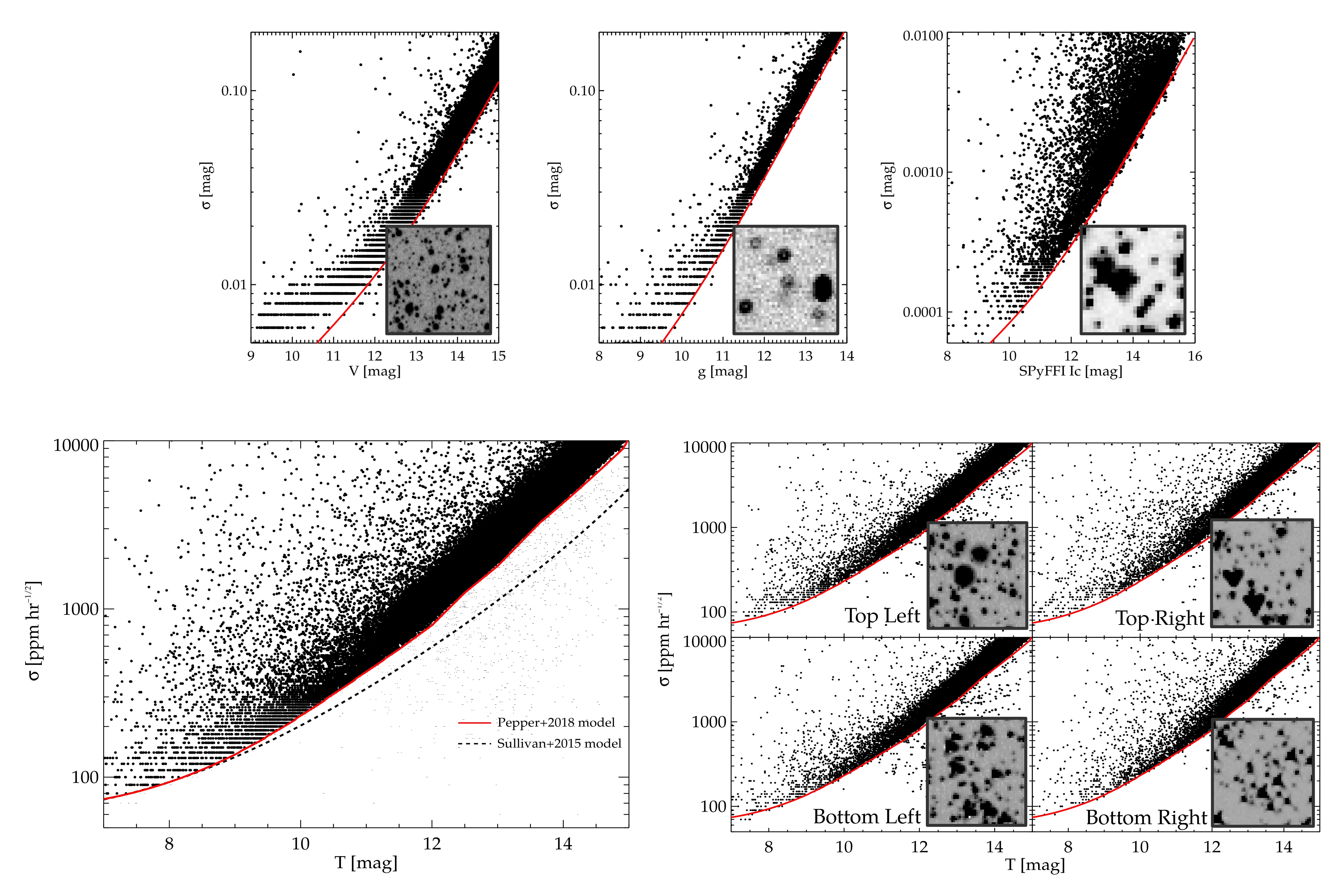}
    \caption{(\textit{Top}:) The achieved rms with the pipeline for 6.4\arcsec/pix \citep{Oelkers:2016b}; the 2009-CSTAR data set with a 15\arcsec/pix with a de-focused PSF \citet{Oelkers2015,Oelkers2016}; and initial testing with the SPyFFI images with $\sim21\arcsec$/pix (Berta-Thompson, 2016 private communication, SPyFFI data).(\textit{Bottom Left}:) The achieved rms for the pipeline described in this work with ETE-6 simulated \textit{TESS} data. The red line shows the expected \textit{TESS} photometric precision from Pepper et al.~(2018, in preparation) and the dashed line shows the expected \textit{TESS} photometric precision from \citet{Sullivan:2015}. A small fraction ($\sim$2\%) of stars, which were over-fit by the detrending routine, are plotted as small black dots under the red line (\textit{Bottom Right}:) The achieved rms as a function of position on the detector. An example PSF is shown in the inset of each panel to demonstrate the changing PSF shape. No discernible difference can be seen between the precision of the light curves as a function of position on the frame.}
    \label{fig:rms}
\end{figure}

The pipeline presented in this paper is a heavily modified version of the pipeline from \citet{Oelkers2015}, which uses the mathematical precepts of \citet{AlardLupton,Miller2008}, and has already been tested on numerous wide-field telescope systems during the past five years. The pipeline has met the nominal precision floor for each system from which data have been fully reduced and published (see Figure~\ref{fig:rms}, top). These systems have ranged in pixel scales from 0.24\arcsec/pix \citep{Diaz:2016} to 15\arcsec/pix \citep{Oelkers2015} for published data, and up to 21\arcsec/pix in private testing with unofficial \textit{TESS} FFIs (Berta-Thompson, 2016 private communication, SPyFFI pipeline).

%Figure 1 - rms values 

The pipeline was originally designed to reduce data from the Chinese Small Telescope ARray (hereafter, CSTAR) \citep{Zhou2010,Oelkers2015}. CSTAR was a small-aperture ($\sim14.5$~cm), wide-field ($\sim$20~deg$^2$) telescope that was deployed to Dome-A in Antarctica during the Antarctic winters of 2008, 2009, and 2010. The telescope collected more than $10^6$ images of nearly continuous time-series photometry of the south celestial pole during each $\sim6$~month observing season. Data reductions of each season identified more than 100 new variable stars and transiting planet candidates \citep{Zhou2010,Wang2011,Wang2013,Oelkers2015,Oelkers2016}.
While CSTAR represented a number of advances in its design and scientific results, the system was not without its limitations. The 2009 observing season had a number of technical issues that greatly stressed the data reduction \citep{Oelkers2015}. The most prominent feature of the data set was the defocused PSF that persisted through all three working cameras, which led \citet{Oelkers2015} to modify the \citet{AlardLupton} and build upon the \citet{Miller2008} DIA routines using a Dirac $\delta$-function kernel to compensate for the changing, highly irregular PSF. 

The current version of the pipeline is composed of 8 routines designed to extract light curves from the FFIs in three steps: (1) background subtraction and image alignment; (2) master frame creation; and (3) image subtraction, fixed aperture photometry, and trend removal. All routines for the pipeline are written in IDL and PYTHON (with the exception of the differencing step, which is written in C with IDL and PYTHON wrappers) to allow increased user flexibility. 

We describe the basic routines below and identify specific IDL and PYTHON routines where appropriate. The pipeline typically fully reduces a single \textit{TESS} FFI (from background subtraction to light curve extraction) in 3~min using Intel Quad-Core Xeon 2.33GHz/2.8GHz processors, and the parameters described in this work.

\subsection{Background Subtraction and Image Alignment \label{subsec:background}}

The fully calibrated FFIs exhibited a low-frequency residual sky background, which is meant to represent the zodiacal background and faint contamination from galaxies and unresolved stars \citep{Stassun:2017}. We applied a residual background subtraction following the approach of \citet{Wang2013, Oelkers2015}. The residual background model is constructed by sampling the sky background every $32\times 32$ pixels over the entire CCD, excluding bad or saturated pixels. A model sky is then fit inside each box and interpolated between all boxes to make a ``thin plate spline" \citep{Duchon1976}. We used the IDL implementation GRID$\_$TPS and the PYTHON implementation of Rbf to make the splines that are then subtracted from every frame. 

Difference imaging requires precise frame alignment in order to produce a proper subtraction. Slight variations caused by improperly aligned frames can contribute to poorly measured fluxes, which will introduce additional dispersion in the extracted light curves. While the \textit{TESS} pointing is expected to be much better than 1~pixel, spacecraft jitter is likely and has been introduced into the simulated FFIs. We use the world coordinate solution (WCS) from the image headers to align the images using the IDL implementation of HASTROM and PYTHON implementation of HCONGRID with cubic-spline interpolation. 

\subsection{Master Frame Combination \label{subsec:master}}

A high-quality master frame is required to preserve the precision of the extracted photometry. Typically this frame is generated by median-combining many individual frames with high SNR and the best seeing, obtained throughout the observing campaign. We created our master frame by median combining all 1,348 images. To save machine memory, images are combined in sets of 50, producing 26 temporary master frames, which are then combined into a single final master frame. 

Next, we identified all 106,399 stars in the TIC with \textit{TESS} magnitude ($T$) $T<16$ that were within the master frame field of view. We then calculated the curve of growth for a variety of aperture sizes, and selected the aperture radius that optimizes the flux calculation for the majority of the stars on the frame. We found this aperture size to be 2.5~pixels.

Fluxes for all stars were then extracted from the master frame using fixed-aperture photometry. We defined the zeropoint offset between the instrumental magnitude scale and the TIC magnitude scale as the median difference between the instrumental magnitude and the TIC $T$ magnitude for all stars. We found the zeropoint to be 
4.825~mag\footnote{We note that for a small number of stars we found a different median zeropoint offset that was $\sim$0.75~mag brighter than the offset above. This suggests a second star of roughly equal brightness is present in the simulated FFIs at the same location, even though only one TIC object exists at that location. This 0.75~mag different offset was found only around some, but not all, bright stars, $T<11$. Therefore, we accepted the zeropoint offset that was consistent with the majority of stars \citep{Stassun:2017}.}.

\subsection{Image Subtraction and Aperture Photometry\label{subsec:dia}}

Typically, DIA routines use an adaptive kernel, $K(x,y)$, defined as the combination of 2 or more Gaussians. While effective at modeling well-defined, circular PSFs, this kernel has difficulty properly fitting other PSF shapes, particularly for highly distorted stars near the edge of the frame \citep{Miller2008}. The \textit{TESS} PSF varies quite significantly across the frame. Therefore, we used a Dirac $\delta$-function kernel to compensate for the non-circular, irregular PSF shape. We defined the kernel as
\begin{equation}
	K(x,y) =  \sum_{\alpha=-w}^w\sum_{\beta=-w}^w c_{\alpha,\beta}(x,y)K_{\alpha, \beta}(u,v)
\end{equation}
\noindent{where $K_{\alpha, \beta}$ is a combination of $(2w+1)^2$ $\delta$-function basis vectors and $K_{0,0}$ is the centered $\delta$~function \citep{Miller2008}. We defined our basis vectors to ensure a constant photometric flux ratio between images \citep{Alard2000, Miller2008}. In the case of $\alpha \neq 0$ and $\beta \neq 0$,}
\begin{equation}
	K_{\alpha, \beta}(u,v) = \delta(u-\alpha,v-\beta) - \delta(u,v) 
\end{equation}
\noindent{while for $\alpha = 0$ and $\beta = 0$,}
\begin{equation}
	K_{0,0}(u,v) = \delta(u,v).
\end{equation}

Stamps are taken around at most 500 bright, isolated stars to solve for the coefficients $c_{\alpha, \beta}(x,y)$ using the least-squares method. We allowed $c_{0,0}(x,y)$ to be spatially variable to compensate for imperfect flat-field corrections. Stars are identified as suitable candidates for stamps provided their measured photometric error was $<50$~mmag and there were no other, brighter stars within 3~pixels of the target star. We found that we could use a $5\times5$~pixel kernel across the frame without appreciably affecting the quality of the subtraction, but significantly reducing the runtime. 

We set the photometric extraction aperture at a 2.5~pixel radius ($52.5\arcsec$) and set the image background to be the value determined in \S~{\ref{subsec:background}}. The differential flux was then combined with the flux from the reference frame, and zero-pointed using the offset described above. Finally, the photometric errors were re-scaled following a methodology similar to \citet{Kaluzny1998}.

\subsection{Light Curve Detrending \label{subsec:detrend}}

Some low-level systematics persisted in the extracted light curves, even with the care taken to properly pre-process and difference the images. While the exact source of these systematics is unknown, the likely explanations include improper image alignment, imperfect stellar stamp selection, minor variations in the calibration process, or a combination of these and other factors. 

We opted to remove (detrend) these systematics from the photometry using an ensemble of light curves to craft a model of non-astrophysical signals. First, for each star, we identified the 1,000 closest stars of similar magnitude ($\sim\pm0.1$~mag). Then we identify a subset of stars which, when subtracted from the light curve, decrease the root-mean-square (rms) variations of the light curve by at least $10\%$. Stars that do not decrease the rms are discarded from the model. If no comparison stars were found to decrease the rms of the target star, then no detrending was applied. If multiple trend stars were found to decrease the rms of the target light curve, then they were median combined into a master model. 

During this process, we noticed that many stars exhibited abrupt changes in magnitude, typically after 48 observations---which is 1~d given the 30-min FFI candence. These changes in magnitude may represent the jitter that was added to the images to simulate mechanical effects. We removed these offsets by subtracting subsequent data points in a given trend model, and identifying observations where the deviation between data points was larger than 5$\sigma$ of the mean deviation between data points. Each subsection of the trend model was then scaled to match the median magnitude of the light curve during this subset. However, if subtracting the scaled trend did not improve the rms by more than $5\%$, the trend during this subsection of the light curve was not scaled. This was done to ensure that true astrophysical signals would not be removed in the detrending routine. 

\begin{figure}[!ht]
    \centering
    \includegraphics[width=\linewidth]{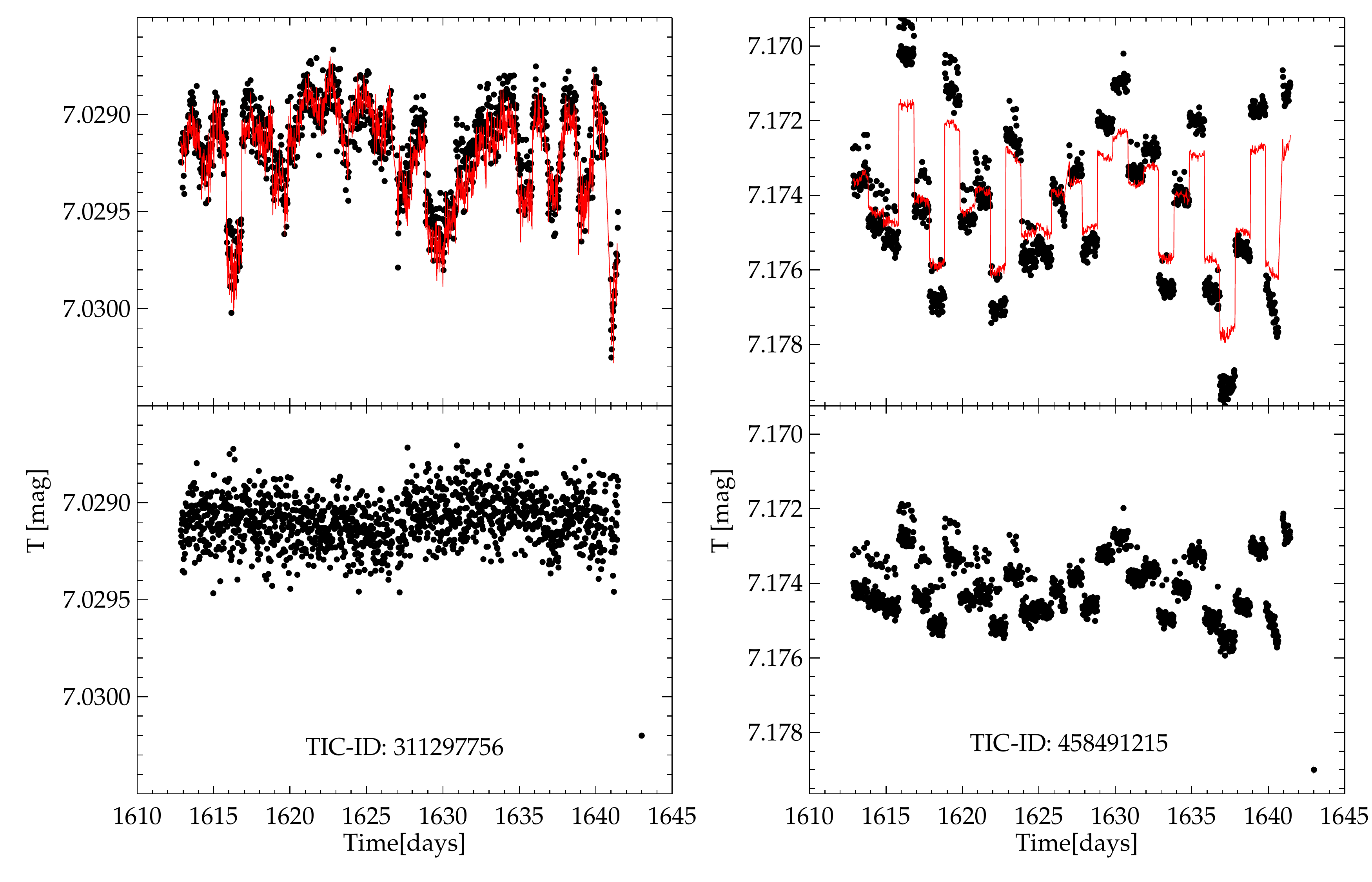}
    \caption{Two light curves from the simulated data set, before (\textit{top}) and after (\textit{bottom}) undergoing the detrending process. Typical photometric uncertainties are shown on the bottom right of each bottom panel. The model trend (red line) is built from a set of stars with similar magnitude that produce at least a $10\%$ improvement in the star's rms. While the star on the left is visibly cleaned by the detrending process, the star on the right still suffers from systematics; we estimate such systematics persist in $\sim$2\% of light curves.}
    \label{fig:detrend}
\end{figure}

Figure~\ref{fig:detrend} shows an example of this detrending and its proficiency at removing non-astrophysical signals. While the detrending routine was proficient at reducing the rms of many stars, there were still stars that failed the detrending process (see Figure~\ref{fig:detrend}, right). We mention these artifact signals as a caution, but note that these events are uncommon in our tests and likely affect $\sim$2\% of stars.

%Figure 2 - Sample Detrending Procedure

Of course, systematics typically vary greatly between systems, and the systematics of the simulated FFIs analyzed here may not fully capture the full systematics of the final, true FFIs after \textit{TESS} launch. Therefore, we propose several methods that users can implement, and which we plan to include with the final release of the FFI data products. These methods include:
\begin{itemize}
    \item Position based detrending: While the pointing of the telescope is expected to be exceptional, it will not be immune to slight corrections for drift. One can use the correlation of time and a star's $x,y$ position to remove trends which appear as the star's centroid shifts from pixel to pixel. This can also be done for sub-pixel movement \citep{Wang2013,Vanderburg:2014}.
    \item Magnitude-based detrending: This method loosely follows the plan we outline above. Stars of similar magnitude can be combined to make a model of variations that are not astrophysical in nature. Typically, two stars of similar magnitude should not vary in identical astrophysical patterns between images, even if they were the same spectral type or variable type. 
    \item Fourier-based detrending: By investigating the power spectrum of the light curve, multiple stars may show similar periods, or aliases of similar periods. These stars could be combined to create a trend pattern, if the period is known \textit{a priori} to be spurious.
\end{itemize}

\section{Results: Pipeline Demonstration with the Simulated Full Frame Images\label{sec:testing}}

\subsection{Performance of the $\delta$ Function Kernel}

The quality of a differenced image can be quantified using stars that cleanly subtract in an image. Specifically, the quality of a differenced frame can be described by assessing the degree to which the deviations left in the differenced frame match the expectations for the noise from the science frame and the master frame \citep{Alard2000}. 

\begin{figure}[!ht]
    \centering
    \includegraphics[width=\linewidth]{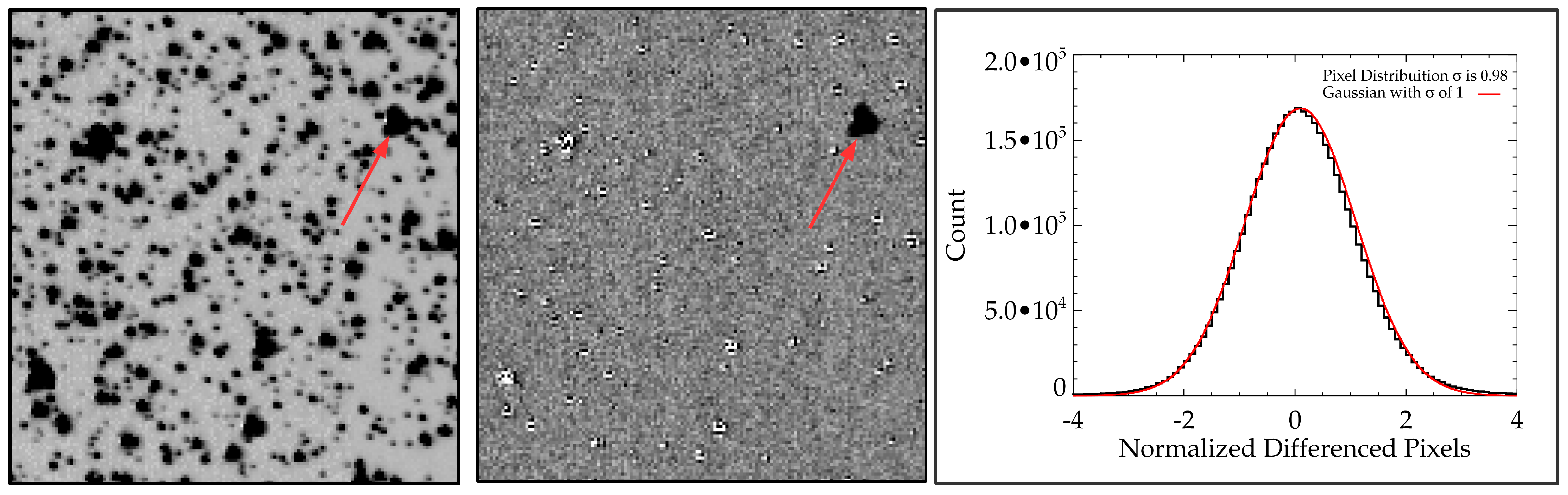}
    \caption{$150\times150$~pixel ($52.5\arcmin\times52.5\arcmin$) cutouts of a typical science frame (\textit{left}) and differenced frame (\textit{center}). The color has been inverted on each frame for clarity. The majority of the stars on the differenced frame either subtract cleanly or show uncorrelated residuals. The red arrows point to an area of correlated positive residuals suggesting a possible variable object. (\textit{Right}:) The normalized histogram of differenced pixel values on the differenced frame (black line) with an over-plotted Gaussian of $\sigma=1$ (red line). The normalized distribution and Gaussian match quite well, suggesting the noise in the differenced frame is meeting the expectation.}
    \label{fig:hist}
\end{figure}

To show that our differenced frames match the expectations for the noise, we select a frame and normalize each pixel value by the combination of the noise from the science frame and the master frame. We defined the expected noise\footnote{This definition differs slightly from the \citep{Alard2000} model, which uses the {\it convolved} master frame noise. We found the two models do not produce significantly different results and therefore accepted the simpler of the two for our analysis.} as: $\delta=\sqrt(I_N+R_N)$, where $I_N$ is the photon counts in the science image and $R_N$ are the photon counts in the master frame. If the subtraction generally matches the expectations from Poisson noise, the normalized pixels should show a Gaussian distribution centered at 0 with a standard deviation of 1. Figure~\ref{fig:hist} shows the histogram of pixel values for a typical differenced frame normalized by $\delta$. The normalized residuals show a mean of 0.122 and a standard deviation of 0.98. We accept this as sufficient evidence that the residuals in the differenced frame is closely matching the expected Poisson noise.

%Figure 3 - Normalized Residuals & Difference Frame

\subsection{Light Curve Precision \label{subsec:noise}}

Next, we compared our resulting light curves to the current models of expected \textit{TESS} precision as a function of stellar magnitude. In particular, we use the original noise model from \citet{Ricker:2014,Sullivan:2015} and the updated noise model from Pepper et al.~(2018, in preparation). These models incorporate shot noise from the star, sky background (in the case of \textit{TESS}, contamination from zodiacal light, unresolved stars, and background galaxies), the read noise of the detector, and a mission-specified noise floor of 60~ppm on 1-hr timescales. Pepper et al.~(2018, in preparation) included an additional estimation of the contamination by nearby stars, which we have excluded from our model because this effect is present in the simulated FFI data through physical contamination, and is manifested in our analysis by stars that deviate from the expected noise floor. These noise models differ quite substantially, particularly in the sky-dominated regime. This is principally because  \citet{Ricker:2014,Sullivan:2015} used an initial optimal-aperture size estimate, while Pepper et al.~(2018, in preparation) uses an updated optimal aperture model based on lab testing of the cameras. We adopt the Pepper et al.~(2018, in preparation) model as a more up-to-date estimation of the photometric precision \footnote{We note that we smoothed the fourth-order estimate of the optimal aperture from Pepper et al.~(2018, in preparation) for our analysis, because we wanted to remove the structure it created in the final model.}, but we include both models for comparison. 

As shown in Figure~\ref{fig:rms} (bottom left), our pipeline very satisfactorily reproduces the expected noise floor of Pepper et al.~(2018, in preparation); the noise floor model matches the lower envelope of stars, representing non-variable objects. Stars above the lower envelope noise floor are interpreted to represent variable objects of various types (see below).

Wide-field imagers typically have a PSF that is largely dependent on the location of the stars on the detector, and the severity of the non-circular PSF shape can be exacerbated near the edges of the frame \citep{Pepper2007}. Since the shape of the PSF is largely position-dependent on the simulated FFIs, we checked $1024\times1024$~pixel subimages in the four CCD corners to see if the precision changed as a function of detector position. As shown in Figure~\ref{fig:rms} (lower right) there is no appreciable difference on any part of the frame.

\subsection{Identifying Variable Stars and Transiting Planet Candidates}

The data release notes for the simulated FFIs indicate that several hundred variable-star and planet-transit signatures were injected into the FFIs to enable checks of signal recoverability. We searched for stars in the simulated data that showed stellar variability and/or transit candidates using the variability metrics of \citep{Oelkers:2018}, a basic Lomb-Scargle periodicity search \citep[LS;]{Lomb,Scargle}, and the box-least-square \citep[BLS;][]{Kovacs2002} algorithm to identify transit-like events.

\begin{figure}[!ht]
    \centering
    \includegraphics[width=\linewidth]{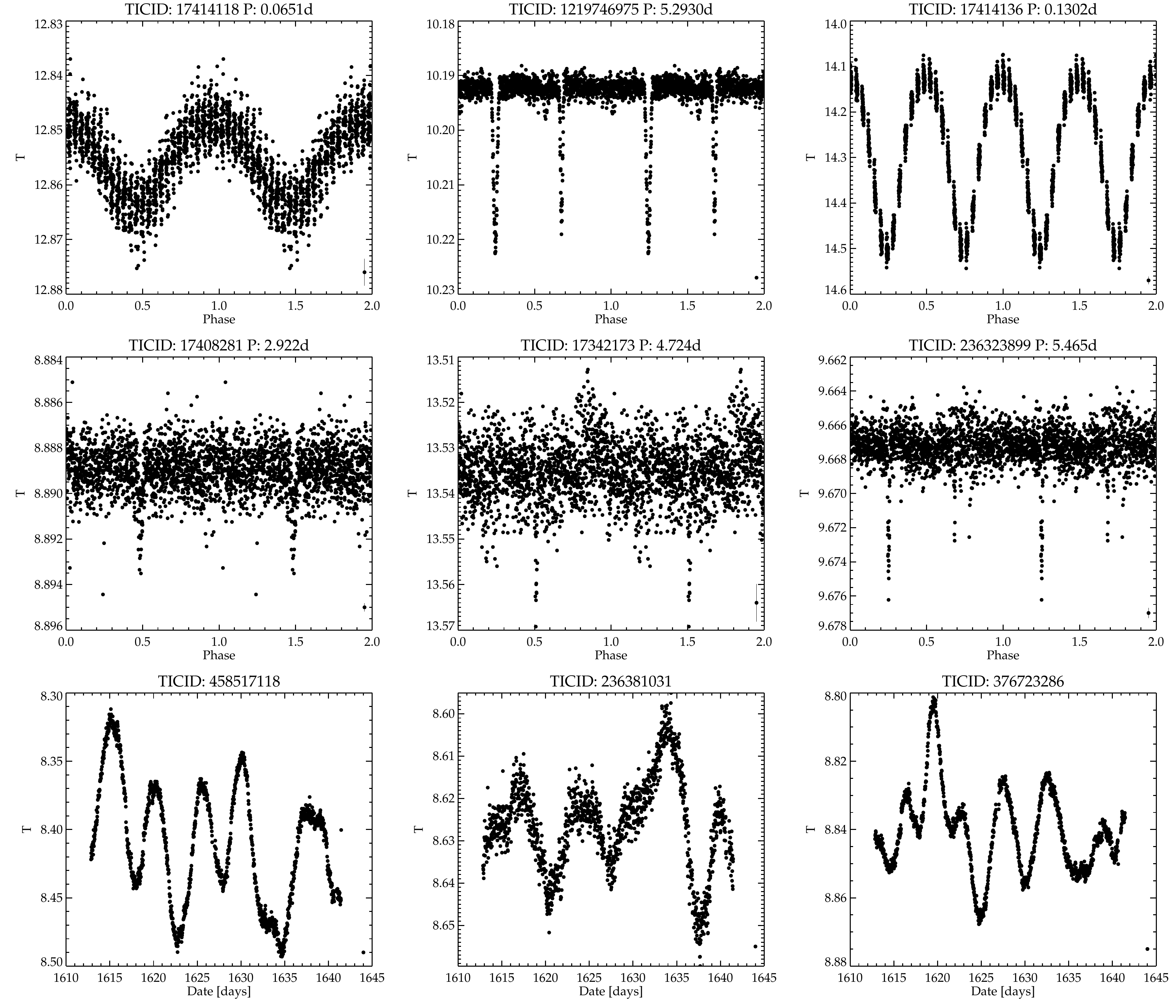}
    \caption{Nine representative light curves from the simulated data set: three likely simulated periodic variable stars (\textit{top row}); three likely simulated transit and detached binary candidates  (\textit{middle row}); and three simulated large amplitude variable stars (\textit{bottom row}). We caution that until we know which stars were indeed injected with variable and transit like signatures some of our identified variable objects may later be determined to be spurious. Phase light curves have been phase-folded and plotted twice for clarity, but not binned. If the recovered period was shown to be an alias of the true period, we phased the light curve on the true period, and not the alias. Representative photometric error bars can be seen at the bottom right of each panel.}
    \label{fig:lc}
\end{figure}

The variability metrics of \citet{Oelkers:2018} were shown to work well to identify stars with large amplitude variability in time-series data. We calculated the metrics \citep[rms, $\Delta90$, Welch-Stetson $J$ and $L$;]{Stetson1996,Wang2013} for every star in the frame, and each star has their specific metric values compared to stars of similar magnitude (for rms and $\Delta90$ metrics) or to the entire field (for $J$ and $L$ metrics). Stars that have metric values larger than a threshold (typically $+2\sigma$ for rms and $\Delta90$; and $+3s\sigma$ for $J$ and $L$) are considered variable. We visually inspected light curves that passed these thresholds and have plotted example stars that appear to have had variability injected into their light curves in Figure~\ref{fig:lc}. The metric values calculated for every star are included in our data release.

We identified periodic signals using two methods. First, we ran a basic LS search \citep{Lomb,Scargle} for periods between 0.01~d and 27~d. We then visually inspected light curves of stars with unique periods, and an SNR for the LS period of $>+3\sigma$ greater than the mean SNR for the entire field. 
Second, we ran a BLS search \citep{Kovacs2002} for periods between 0.3~d and 27~d with 10000 frequency steps and 100 phase bins. We selected stars for visual inspection if their signal-detection-efficiency (SDE) values were larger than $>+2\sigma$ of the mean SDE value for the field, and if the target period has less than 50 stars with an identical period. Typically, heuristic cutoffs will be identified using the data from the entire frame to identify the most likely astrophysical events. However, because events were injected into a small number of stars ($<15,000$ over all 16 CCDs), and we currently do not have a list of all objects with injected signals, we settled for a basic cut on period and visual identification for quality testing purposes. In the future, we plan to detect events using a methodology similar to previous work \citep{Wang2011,Wang2013,Rodriguez:2017tau,Ansdell:2018,Oelkers:2018}. We identified several possible planet-transit candidates, and have plotted two such candidates in Figure~\ref{fig:lc}. 

Using the metric above, we identified 2,275 stars with a possible variable signal, 64 stars with significant LS periods, and 64 stars with significant BLS periods based on the above metrics. Some objects found with the variability search were also identified as poorly detrended stars. We caution that until we know for certain which stars were actually injected with the variable and transit-like signals, some of the identified events could be spurious. In any case, we include all calculated metrics in our data release portal (see below) to allow users to experiment with the data using different significance thresholds.

%Figure 4 - Sample Light curve

\section{Summary \label{sec:summary}}

We have presented a data reduction pipeline that has been adapted for \textit{TESS} FFIs. The pipeline includes routines for background subtraction, image alignment, master frame combination, difference imaging, aperture photometry, and magnitude-based trend removal. The final precision of the produced light curves has been shown to match the expected photometric precision of the detector. Finally, a variety of injected astrophysical variables and transit signatures have been identified.

The full pipeline described in this paper is currently available for download through a GITHUB repository at the URL \url{https://github.com/ryanoelkers/DIA/}. We have tested the routines on multiple computers with multiple data sets, and we believe the code can be readily used as is or adapted. Some external libraries are required, and we list them with the README file at the repository. The pipeline is currently in a `version zero' state, and users should expect the pipeline to be updated and improved prior to the public release of \textit{TESS} data from the first observed sector, expected for late 2018. 

We encourage adaptation of the code, and we encourage users to cite this work as well as \citet{Alard2000,AlardLupton,Miller2008,Oelkers2015} given the large contributions of those works to the formulation of the pipeline presented here. 

We plan to use our pipeline to reduce \textit{TESS} FFIs and we will release all data products to the public on a rapid timescale\footnote{Currently, we estimate this will be approximately 1--2 weeks after a \textit{TESS} public data release given our current computing resources.}. All data products will be released through the \textit{Filtergraph} visualization portal at the URL \url{https://filtergraph.com/tess_ffi}. Users will be able to use the portal to access TIC information for the extracted stars, light curve data files, image files of light curves, and links to Simbad and Aladin for each star. We will also provide basic variability information calculated for each light curve using the metrics from \citep{Oelkers:2018}, and periodicity information using a Lomb-Scargle analysis \citep{Lomb,Scargle}, as well as a best fit box-least-square period \citep{Kovacs2002}. We encourage readers to visit the \textit{Filtergraph} URL even in advance of the first sector data release to see an example of the data release using the data from this work, and welcome suggestions for how the data release can be improved in the future.

\acknowledgements

The authors would like the \textit{TESS} Science Processing Operations Center for making the simulated images available for the community to download and experiment with. R.J.O. would like to thank the Stassun group for their useful critiques, and suggestions which greatly improved the quality of this manuscript. This work has made extensive use of the \textit{Filtergraph} data visualization service \citep{Burger:2013} at \url{http://filtergraph.com}. This research has made use of the VizieR catalogue access tool, CDS, Strasbourg, France. This work was conducted in part using the resources of the Advanced Computing Center for Research and Education at Vanderbilt University, Nashville.  We acknowledge use of the ADS bibliographic service. This work has made use of the TIC and CTL, through the TESS Science Office's target selection working group (architects K.~Stassun, J.~Pepper, N.~De~Lee, M.~Paegert, R.~Oelkers). 

\bibliographystyle{apj}
\bibliography{references}

\end{document}